# Multi-Scale Temporal Analysis for Failure Prediction in Energy Systems

Anh Le, PhD Student, North Dakota State University

Phat K. Huynh, PhD, North Carolina A&T State University

Om P. Yadav, PhD, North Carolina A&T State University

Chau Le, PhD, University of North Carolina at Charlotte

Harun Pirim, PhD, North Dakota State University

Trung Q. Le, PhD, University of South Florida

## SUMMARY & CONCLUSIONS

Many previous studies on power networks focus on identifying disturbances but they often miss how the system behaves across different time scales. Many existing models struggle to predict these systems' complex, nonlinear behavior of these systems during extreme weather conditions. This study proposes a multi-scale system temporal analysis for failure prediction in energy systems, focusing on predicting failures in electric power networks using Phasor Measurement Unit (PMU) data. The model integrates multi-scale analysis with advanced machine learning techniques to capture both short-term and long-term behavior. By analyzing disturbances across multiple temporal scales, we aim to improve failure prediction accuracy in complex energy systems. One of the main challenges with PMU data is the lack of labeled states, despite the availability of disturbance times logged as failure records. This limitation makes distinguishing between normal, pre-disturbance, and post-disturbance conditions difficult. We apply multi-scale system dynamics modeling to address this issue and extract key predictive features from PMU data for failure prediction. First, multiple domain features were extracted from the PMU data from time series data. Second, a multi-scale windowing approach was applied, using window sizes of 30s, 60s, and 180s to capture both short-term and long-term system dynamics. This multi-scale analysis allowed the model to detect patterns across different time scales. Third, feature selection was performed using Recursive Feature Elimination (RFE) to reduce the dimensionality of the dataset and identify the most important features. We selected RFE as the best feature selection for its ability to systematically eliminate less important features, thus improving model interpretability and reducing complexity. This step was crucial for identifying the most relevant features for predicting disturbances. Fourth, these features were used to train multiple machine-learning models. Our main contributions include:

1. Identifying significant features across multi-scale windows.
2. Demonstrating LightGBM as the best-performing model, achieving 0.896 for precision with multi-scale windows.
3. Showing that multi-scale window sizes (30s, 60s, 180s) significantly improved model performance by capturing both short-term and long-term trends, outperforming single-window models (0.841 score).

Our current work focuses on weather-related failure modes, with future research extending this approach to other failure modes, equipment failure, and lightning events.

## 1 INTRODUCTION AND LITTERATURE REVIEW

Smart grids, the future of power grids, have offered reliable, resilient, and efficient solutions for power system management[1]. These grids can be categorized into open grids and underground grids. Open grids are directly exposed to weather elements, making them more prone to failures caused by weather disturbances, lightning, fires, and even human error. Underground grids are protected from many of these external factors, but they face issues like insulation degradation due to moisture and thermal stresses [2]. Underground grid failure detection systems should identify overloads, locate faults precisely, and isolate defective sections [3].

This research focuses on open grid systems, which are highly vulnerable to frequent weather disturbances. Weather-induced failures cause widespread power outages and can also trigger cascading failures. These failures may lead to equipment overloads, transformer damage, and disruptions to critical services, such as hospitals and emergency systems. Despite advancements in grid technology, current predictive models are insufficient for accurately forecasting these failures under extreme weather conditions. This highlights a significant research gap in failure prediction and resilience enhancement for open grid systems.

Over the past few decades, various methodologies have been proposed for anomaly detection and prediction in power systems. These approaches can be broadly categorized into stochastic and machine learning/deep learning (ML/DL) models. Stochastic models have become increasingly popular for power system fault diagnosis and anomaly detection. In 2020, robust signal recovery, event outlier detection, and stochastic subspace selection with principal component analysis (R-PCA) were used to identify anomalies in synchrophasor data [4]. However, this method was unable to differentiate between event-induced and spurious outliers, often leading to incorrect signal reconstruction. In 2023, the authors advanced this work with the supervised multi-residual generation learning (SMGL) model, which diagnoses single-phase-to-ground faults by integrating active distribution system protection as a Markov decision process [5]. Nevertheless, the SMGL model's focus on single-phase-to-ground faults and its robustness in measuring noise and PMU inaccuracies.

The advent of ML/DL techniques has revolutionized anomaly detection and prediction in power systems. Various state-of-the-art models, including Convolutional Neural Networks (CNNs) [6, 7] have been applied to capture the complex patterns in PMU data. These models have shown superior performance compared to traditional approaches. Among these models, CNNs have shown promise in anomaly detection and prediction tasks. For instance, in 2021, 1D-CNNs [6] were used for real-time short-term voltage stability assessment, based on a limited number of PMU voltage samples. Further advancements in CNNs were made in 2023 by transforming several types of phasor time series measured at various locations into generalized multichannel images [7]. The CNNs were trained to detect suspicious patterns in these images.

Recent studies show that multi-scale methods work well in different fields; however, failure prediction is still limited. For example, Conv1D models have performed well in tasks like crop classification [8]. However, their ability to capture both short-term and long-term trends in energy systems needs more research. Current event detection methods in large power systems often focus on single or specific event types [9, 10]. Most of these approaches relied on limited data like frequency measurements [9]. It did not use other data types that could help better detect failure. The multivariate singular spectrum analysis model mainly focused on simulated normal system behavior instead of real-world energy environments [11]. Other techniques, such as kernel density estimation and cyber-physical event detection models, have their limits when dealing with complex or multiple events simultaneously [12]. While extended multi-scale principal component analysis has proven that multi-scale methods are better than single scale for detecting anomalies, it has not been fully applied to energy system failure prediction [13]. Multi-scale techniques from other fields, like texture analysis in biomedical research, suggest ways to study multi-scale dynamics in energy systems [14]. There is a clear need to develop strong multi-scale models for predicting failures in energy systems. These models should handle complex systems, adapt to changes, and avoid frequent retraining.

We propose a multi-scale temporal analysis to predict network failures in energy systems, focusing on weather-related disturbances. The model captures both short-term and long-term temporal dependencies in PMU data, enabling a better representation of temporal dynamic relationships. Our approach involves crafting features that reflect multi-scale dynamics through time and frequency domain analysis, followed by Recursive Feature Elimination (RFE) to identify key features and manage high-dimensional data. LightGBM was identified as the best-performing model, effectively learning complex nonlinear patterns and outperforming other machine-learning models. We applied our model to a dataset of 446 instances across different disturbance states. Subsequently, we demonstrated the value of combining multi-scale temporal with machine-learning model to improve failure prediction accuracy.

## 2 METHODOLOGY

This research focuses on extreme weather as the primary failure mode in open grids. While there are many types of grid failures, we selected weather-related disturbances for this exploratory study due to their prevalence and significant impact on the system. Other failure modes, such as lightning strikes, alternating current equipment failures, or circuit malfunctions, will be explored in future studies. Our approach comprises three main components: data preparation, feature engineering, and machine-learning modeling.

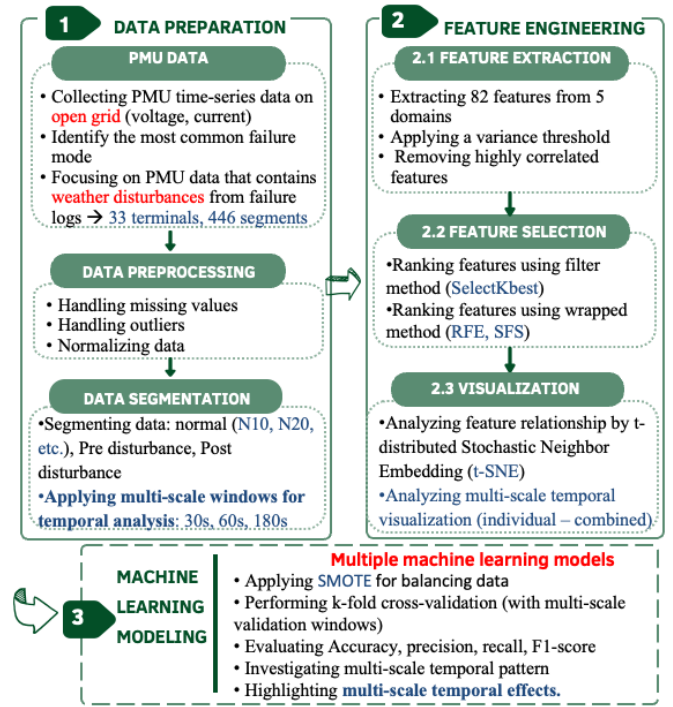

*Figure 1. Methodology framework*

### 2.1 Data preparation

From January 2020 to August 2022, high-resolution data was collected from 536 Phasor Measurement Units (PMUs)

strategically installed across regional power stations in Oklahoma. These PMUs continuously monitored and recorded real-time voltage and current magnitudes at a sampling rate of 30 Hz. Our study focused on weather-related disturbances, such as storms, which cause many of the power grid failures. To analyze the impact of these disturbances comprehensively, data were gathered on three separate days: July 10th, October 27th, and October 28th, 2020. Each of these days experienced multiple weather disturbances, offering a range of conditions for analysis.

Starting with 446 segments for both voltage and current (892 total), outliers were removed using a threshold of the mean plus three times the standard deviation, resulting in 802 entries for our analysis. Out of the 802 total entries, 495 (Nor) represent normal states, 153 (Pre) represent pre-disturbance states, and 153 (Post) represent post-disturbance states. This situation means imbalance. Min-Max normalization was applied to scale features, ensuring comparability across features. The data were segmented into time windows based on disturbance events from the historical failure log. The normal class consists of N50, N40, N30, N20, and N10 (representing 50, 40, 30, 20, and 10 minutes before recorded disturbance), Pre (immediately before), and Post (immediately after). Segmentations allowed for detecting of early warning signs that may appear at different times. Multi-scale analysis was performed using window sizes of 30s, 60s, and 180s for each segment to capture both short-term and long-term dynamics (*see Figure 2*). We adopted 30-second windows following the findings in these references [9,15], which showed its effectiveness in capturing critical short-term transitions that precede disturbances.

*2.2 Feature engineering*

Feature extraction: A comprehensive set of features was extracted to capture the power system's complex dynamics *(see Table 1)*. These features include time-domain, frequency-domain, and system dynamics-specific measures, such as covariance and detrended fluctuation analysis. They capture time-based relationships and long-term patterns, which are important for predicting failures in power systems.

*Table 1: Description of the extracted features.*

| Domain | Features | Description |
|---|---|---|
| Time | Mean, Variance, Skewness, Kurtosis | These features represent central tendency, asymmetry, and shape of distribution. |
| Frequency | FFT Coefficients, Spectral Entropy, Spectral Centroid | These features reveal hidden patterns and dominant frequency. |
| Information Theory | Shannon Entropy, Permutation Entropy, Sample Entropy | These measures catch changes in signal patterns. These help spot early signs of power system problems. |
| Energy | Energy, Root Mean Square, Peak-to-Peak Amplitude | These features measure power levels and sudden changes in the signal. |
| System Dynamics | Covariance Matrix, Detrended Fluctuation Analysis | These features identify temporal correlations and long-range dependencies in system behavior. |

Feature Selection: Among the methods tested, we identified Recursive Feature Elimination (RFE) as the most effective method for feature selection. RFE simplified the model by reducing irrelevant features, making it more efficient and reliable. Moreover, the visualization of top 20 features selected by RFE was significantly clearer.

*2.3 Machine Learning Model Deployment*

In this study, the data was imbalanced because the Pre- and Post-disturbance have fewer instances than the normal class. Synthetic Minority Over-sampling Technique (SMOTE) was applied to handle imbalanced datasets to increase the number of minority classes. The dataset, split into training and testing sets with a 90:10 ratio, ensured robust training on 90% of the data and validation on the remaining 10%. SMOTE was applied to oversample the pre-disturbance (153 instances) and post-disturbance data (153 instances), resulting in 495 instances (equal to the normal class). We then used 10-fold cross-validation (K=10) to evaluate the performance of models. We divided the data into ten folds, so most folds had 49 instances per class, while a few had 50. The number 49.5 is just an average and does not mean we used fractions of instances. This method kept the class distribution balanced while ensuring fair evaluation.

To find the best machine-learning model, we applied multiple models such as Random Forest, XGBoost, LightGBM, Decision trees, etc. Evaluation metrics included precision, recall, F1-score, and accuracy to give a balanced view of model performance. These models were trained using a carefully selected set of features. We divided our test data into different segments based on periods. Afterward, we applied our trained model to this segmented test dataset. Performance evaluation was conducted using accuracy, precision, recall, and F1-score metrics to assess the model's performance in the electric power network. They reflect how effectively it can detect and classify the various operational states within the power network's health monitoring system.

## 3 RESULTS AND DISCUSSION

This section presents the outcomes of our exploratory data analysis, feature extraction, and model training. Our PMU data analysis helps us to understand the dynamic behaviors of electric power networks under different operating conditions.

## 3.1 Exploratory data analyses

Initially, to comprehend the complex relationships within the PMU data, we conducted exploratory data analyses to distinguish between normal, pre-disturbance, and post-disturbance states. Our strategy involved extracting diverse features from time series data with multiple domains. This comprehensive approach aimed to study complex system dynamics and distinguish between normal and disturbed states.

*Figure 2* illustrates the voltage magnitude, with a red dashed line marking disturbance times as logged in our failure records. In the context of our study, disturbances refer to events that disrupt the normal operation of the power grid. Specifically, we examined weather-related disturbances based on the failure log.

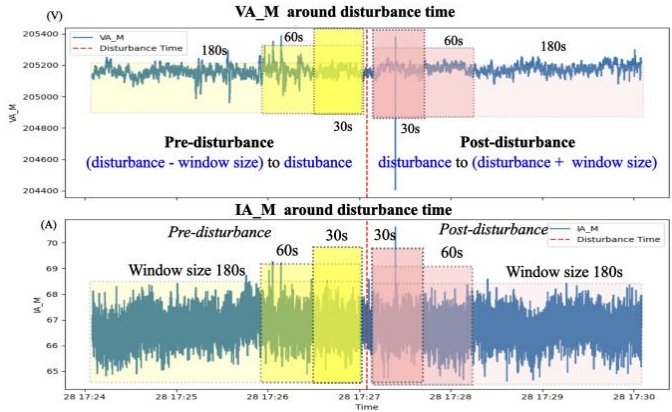

*Figure 2. Voltage magnitude and current magnitude at terminal 112. The red dashed line represents disturbance time in the failure log.*

### 3.2 Feature extraction and selection

In our study, the best feature selection method, RFE, was used. The method was used to identify the top 10, 15, and 20 most significant features from an initial set of 82 features across various domains (Table 1).

*Table 2. Importance coefficient of top 5 features (RFE)*

| Feature(30s) | | Feature (60s) | |
|---|---|---|---|
| Name | Score | Name | Score |
| dfa_VA_M | 0.116 | s_bandw_IA_M | 0.116 |
| w_entropy_VA_M | 0.116 | mean_IA_M | 0.109 |
| cov_IA_M_2 | 0.110 | skewness_VA_M | 0.107 |
| s_entropy_VA_M | 0.010 | dfa_VA_M | 0.102 |
| cov_IA_M_1 | 0.010 | std_VA_M_0 | 0.010 |
| Feature(180s) | | Feature (Total) | |
| Name | Score | Name | Score |
| cov_IA_M_3 | 0.113 | rms_IA_M | 0.130 |
| m_sp_std_IA_M | 0.104 | mean_IA_M | 0.127 |
| mean_IA_M | 0.103 | mean_IA_M_0 | 0.123 |
| mean_IA_M_0 | 0.101 | rms_VA_M | 0.097 |
| dfa_IA_M | 0.101 | std_VA_M_3 | 0.092 |

*Table 2* shows that in the 30s window, the dominance of detrended fluctuation analysis of voltage magnitude (dfa_VA_M) and wavelet entropy of voltage magnitude (w_entropy_VA_M) indicates the importance of detecting rapid changes. In the 60s window, both statistical domains, such as the mean of current magnitude (mean_IA_M) and skewness of voltage magnitude (skewness_VA_M and information theory-based features such as spectral bandwidth of current magnitude (s_bandw_IA_M), are significant. Therefore, there is a need for both in medium-term analysis. In the 180s window, the covariance of current magnitude (cov_IA_M_3) and magnitude spectrum standard deviation of current magnitude (m_sp_std_IA_M) become more important. This suggests that longer windows capture broader patterns. The root mean square of current magnitude (rms_IA_M) and mean of current magnitude (mean_IA_M) remain consistently important across all windows, showing their stability. This suggests that a combination of short- and long-term features may be best for prediction.

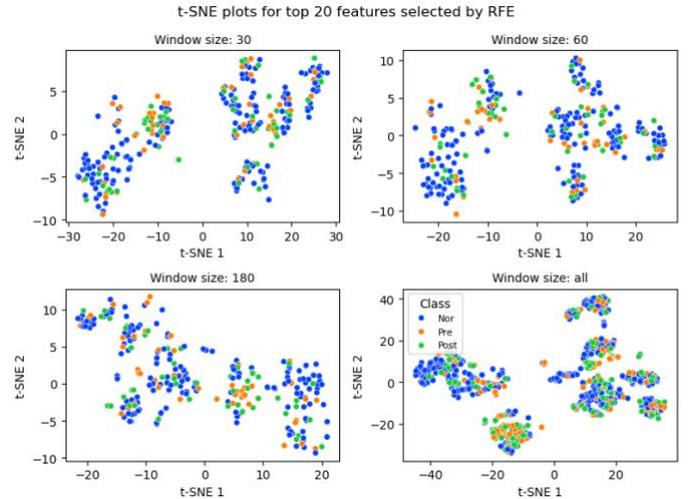

*Figure 3. tSNE Visualization for data by multiscale window size*

*Figure 3* shows that the t-SNE visualizations separate the Normal class from Pre and Post disturbance classes, especially in larger time windows (60s and 180s). This means that feature engineering can reliably detect normal conditions. However, there is an overlap between the Pre and Post classes in all window sizes. This suggests the selected features may not clearly distinguish between pre- and post-disturbance. Since the pre- and post-disturbance classes show a transition, Kulback-Leibler divergence could be explored to highlight gradual changes over time. The visualization using all windows shows the best separation, which supports the multiscale features for better class distinction. However, the overlap between pre- and Post classes suggests that more work is needed to improve the feature selection.

### 3.3 Machine learning model results

In this section, we used confusion matrix to evaluate the performance of multiple machine-learning models. By comparing different models, we identified the most effective approach for capturing the system's complex dynamics *(Table 3)*.

*Table 3* indicates that LightGBM consistently outperforms the other models across all cases, with the highest precision, recall, and F1 scores. It also balances both false positives and false negatives. **Compared to individual window sizes with around 82%, multiscale windows (30s, 60s, 180s) significantly improve model performance to 89.6%**. XGBoost excels in the Top 20 with window size 60s, indicating its strength in capturing medium-term patterns, though it still falls slightly behind LightGBM overall. While Random Forest remains competitive, especially in the shorter time windows, it tends to trail behind LightGBM and XGBoost, making it a solid but less optimal choice because it reduces overfitting by averaging multiple decision trees.

*Table 3. The performance of 3 best machine-learning model*

| Case | Metric | RF | | XGBoost | | LightGBM | |
|---|---|---|---|---|---|---|---|
| | | Mean | Std | Mean | Std | Mean | Std |
| Top 20 30, 60, 180s | Accuracy | 0.855 | 0.025 | 0.880 | 0.024 | 0.895 | 0.028 |
| | Precision | 0.875 | 0.025 | 0.882 | 0.024 | 0.896 | 0.029 |
| | Recall | 0.873 | 0.025 | 0.881 | 0.024 | 0.895 | 0.028 |
| | F1 Score | 0.872 | 0.025 | 0.881 | 0.024 | 0.895 | 0.028 |
| Top 20 30s | Accuracy | 0.806 | 0.053 | 0.802 | 0.078 | 0.816 | 0.050 |
| | Precision | 0.816 | 0.050 | 0.810 | 0.073 | 0.826 | 0.047 |
| | Recall | 0.806 | 0.053 | 0.802 | 0.077 | 0.816 | 0.050 |
| | F1 Score | 0.804 | 0.053 | 0.800 | 0.078 | 0.814 | 0.051 |
| Top 20 60s | Accuracy | 0.812 | 0.050 | 0.848 | 0.061 | 0.837 | 0.062 |
| | Precision | 0.820 | 0.046 | 0.853 | 0.060 | 0.841 | 0.062 |
| | Recall | 0.812 | 0.050 | 0.849 | 0.061 | 0.837 | 0.062 |
| | F1 Score | 0.810 | 0.051 | 0.848 | 0.061 | 0.837 | 0.063 |
| Top 20 180s | Accuracy | 0.814 | 0.056 | 0.804 | 0.048 | 0.826 | 0.031 |
| | Precision | 0.823 | 0.052 | 0.813 | 0.046 | 0.841 | 0.031 |
| | Recall | 0.814 | 0.056 | 0.804 | 0.048 | 0.836 | 0.031 |
| | F1 Score | 0.814 | 0.056 | 0.805 | 0.048 | 0.836 | 0.032 |

Besides the results in *Table 3*, the study also explored other machine-learning models, such as Gradient Boosting, Super Vector Machine, K-Nearest Neighbors, Linear Discriminant Analysis (LDA), and Logistic Regression. However, these models underperformed significantly compared to LightGBM, XGBoost, and Random Forest. Because they struggle with non-linear relationships. Simpler models such as Logistic Regression and LDA struggled to capture the complexity of the data, with an accuracy of less than 60%. The other improves slightly, but their performance is less than 80%.

*Table 4. Confusion matrix of the best model – LightGBM with all window size*

| | Nor | Pre | Post | Total |
|---|---|---|---|---|
| **Nor** | 90.71% | 4.44% | 4.85% | 100% |
| **Pre** | 3.43% | 89.70% | 6.87% | 100% |
| **Post** | 5.25% | 6.67% | 88.08% | 100% |
| **Total** | 100% | 100% | 100% | |

*Table 4* shows the confusion matrix for the top 20 features in the 30s, 60s, and 180s window sizes using the LightGBM model. It demonstrates robust performance across all three states (Nor, Pre, Post), especially in normal (Nor) and pre-disturbance (Pre). Misclassification rates are relatively low. However, there is some confusion between pre-disturbance (Pre) and post-disturbance (Post) instances. This should be refined in the model to better capture the transition between these states. The high precision and recall for normal and pre-event states make LightGBM well-suited for applications.

## 4 CONCLUSION

The study indicates the most significant features with different window sizes. It also demonstrates the robust approach in feature engineering using t-SNE visualization. It effectively detects normal classes. However, there is still overlap between the pre- and post-classes across all window sizes. This implies that the feature engineering does not fully differentiate pre-disturbance from post-disturbance states. Multiscale window size shows the best separation, but more improvement is needed to better distinguish the classes. This aligned with the performance in the machine-learning model. LightGBM- the best model faces challenges when classifying pre- and post-disturbances. LightGBM with built-in feature importance achieves the highest accuracy (0.896 ± 0.028 with 20 features), outperforming Random Forest, XGBoost, and other models. LightGBM builds deeper trees faster than XGBoost and RF, helping it capture complex feature relationships, even in smaller datasets. Future studies could apply this system approach to different failure modes, such as circuit failures or lightning strikes, and improve feature engineering to enhance the performance of the model.

*ACKNOWLEDGEMENTS*

This research is partially supported by National Science Foundation (NSF) EPSCoR RII Track-2 Program under the grant number OIA-2119691 at North Dakota State University.

BIOGRAPHIES

Anh Le, PhD Student
Department of Civil Engineering North Dakota State University
e-mail: ducanh.le@ndsu.edu
Anh Le is currently a PhD Student of the Department of Civil Engineering at North Dakota State University. His research interests focus on advanced deep learning in smart energy systems, computational fluid dynamics.

Phat K. Huynh, PhD
Department of Industrial and Systems Engineering
North Carolina A&T State University
e-mail: pkhuynh@ncat.edu
Dr. Phat K. Huynh is currently an Assistant Professor of the Department of Industrial and Systems Engineering at North Carolina A&T State University. His work focuses on complex systems modeling and predictive analytics in healthcare.

Om Prakash Yadav, PhD
Department of Industrial and Systems Engineering
North Carolina A&T State University
e-mail: oyadav@ncat.edu
Dr. Om Prakash Yadav is currently a Professor and Chair of the Department of Industrial and Systems Engineering at North Carolina A&T State University. His research interests include reliability modeling and analysis, risk assessment, robust design optimization, and manufacturing systems analysis.

Chau Le, PhD
Department of Engineering Technology & Construction Management
University of North Carolina at Charlotte
e-mail: cle12@charlotte.edu
Dr. Chau Le is currently an Assistant Professor in the Department of Engineering Technology & Construction Management at University of North Carolina at Charlotte. His research interests include human safety and health, construction and infrastructure management, and artificial intelligence.

Harun Pirim, PhD
Department of Industrial and Manufacturing Engineering
North Dakota State University
e-mail: harun.pirim@ndsu.edu
Dr. Harun Pirim is currently an Assistant Professor of the Department of Industrial and Manufacturing Engineering at North Dakota State University. His research interests focus on network science, mathematical programming, and machine learning in biological, social, and decision sciences.

Trung (Tim) Q. Le, PhD
Department of Industrial & Management Systems Engineering
University of South Florida
e-mail: tqle@usf.edu
Dr. Trung (Tim) Q. Le is currently an Assistant Professor of the Industrial and Management Systems Engineering Department at the University of South Florida. His research focuses in 3 main directions: 1) data-driven and sensor-based modeling, 2) medical device manufacturing and bio-signal processing, and 3) predictive analytics for personalized healthcare.